\documentclass[aps,prd,superscriptaddress,long,notitlepage,balancelastpage,nofootinbib,floatfix,twocolumn]{revtex4-1}
\pdfoutput=1
\usepackage{amsmath,mathtools,amssymb,amsthm,amsxtra,overpic,bbm,epsfig,subfigure,url,bm}
\usepackage{hyperref}
\usepackage{mathrsfs}
\usepackage{color,xcolor}
\usepackage{comment}
\usepackage{float}
\usepackage{enumitem}
\usepackage{slashed}
\usepackage{multirow}
\usepackage[normalem]{ulem}
\definecolor{nicered}{rgb}{0.5,0.,0.}
\definecolor{nicegreen}{rgb}{0.,0.5,0.}
\definecolor{niceblue}{rgb}{0.,0.,0.5}
\hypersetup{
	colorlinks=true,
	linkcolor=black,
	filecolor=nicegreen,      
	urlcolor=niceblue,
	citecolor=nicered,
}

\makeatletter
\newcommand*{\balancecolsandclearpage}{%
	\close@column@grid
	\cleardoublepage
	\twocolumngrid
}
\makeatother
\begin{document}
%%%%%%%%%%%%%%%%%%%%%%%%%%%%%%

\title{\vspace{1cm} \Large 
Scalar nonstandard neutrino interactions in galactic supernovae 
}

%%%%%%%%%%%%%%%%%%%%%%%%%%%%%%    
\author{Bhaskar Dutta}
\email{dutta@tamu.edu}
\affiliation{Department of Physics and Astronomy, Mitchell Institute for Fundamental Physics and Astronomy, Texas A\&M University, College Station, Texas 77843, USA}
\author{Aparajitha Karthikeyan}
\email{aparajitha\_96@tamu.edu}%
\affiliation{Department of Physics and Astronomy, Mitchell Institute for Fundamental Physics and Astronomy, Texas A\&M University, College Station, Texas 77843, USA}
\author{Nityasa Mishra}
\email{nityasa\_mishra@tamu.edu}%
\affiliation{Department of Physics and Astronomy, Mitchell Institute for Fundamental Physics and Astronomy, Texas A\&M University, College Station, Texas 77843, USA}
\author{Yago Porto}
\email{yago.porto@ufabc.edu.br}%
\affiliation{Centro de Ciências Naturais e Humanas, Universidade Federal do ABC, 09210-170, Santo André, SP, Brazil}
\affiliation{{Instituto de F{\'i}sica Gleb Wataghin, Universidade Estadual de Campinas, 13083-859, Campinas, SP, Brazil}}
\author{Louis E. Strigari}%
\email{strigari@tamu.edu}%
\affiliation{Department of Physics and Astronomy, Mitchell Institute for Fundamental Physics and Astronomy, Texas A\&M University, College Station, Texas 77843, USA}
%%%%%%%%%%%%%%%%%%%%%%%%%%%%%%

%%%%%%%%%%%%%%%%%%%%%%%%%%%%%%%%%%%%%%%%%%%
%%%%%%%%%%%%%%%%%%%%%%%%%%%%%%%%%%%%%%%%%%%
\begin{abstract}
We analyze the prospects for studying scalar nonstandard interactions (SNSI) using the neutrino burst from a galactic supernova (SN). SNSI modifies the resonant flavor conversion and, correspondingly, the neutronization burst signal, and may be identifiable in future multi-tonne-scale experiments, such as Deep Underground Neutrino Experiment (DUNE). Notably, in the normal mass ordering, SNSI can lead to an anomalous (re)appearance of the neutronization-burst $\nu_e$ peak at Earth, a feature otherwise strongly suppressed in the Standard Model. We show that, in the presence of SNSI, neutrinos propagating out of the dense SN environment acquire a density-squared-dependent contribution to their mass-squared differences, which in turn modifies the energy levels of the neutrino matter eigenstates. This density-squared effect is absent in less dense environments such as the Earth or the Sun and is precisely what enables SN neutrinos to probe SNSI parameters up to 4 orders of magnitude below the values accessible to solar or terrestrial neutrino sources.
%%%%%%%%%%%%%%%%%%%%%%%%%%%%%%%%%%%%%%%%%%%
%%%%%%%%%%%%%%%%%%%%%%%%%%%%%%%%%%%%%%%%%%%

\noindent 
\end{abstract}
\maketitle
%%%%%%%%%%%%%%%%%%%%%%%%%%%%%%%%%%%%%%%%%%%
%%%%%%%%%%%%%%%%%%%%%%%%%%%%%%%%%%%%%%%%%%%
\section{Introduction} 

The observation of neutrinos from SN1987A~\cite{Kamiokande-II:1987idp, Bionta:1987qt, Alekseev:1988gp} confirmed the basic picture of core-collapse SNe, in which more than $99\%$ of the released gravitational binding energy is emitted in the form of neutrinos of all flavors~\cite{Mirizzi:2015eza, Janka:2017vlw}. Since then, neutrino detectors have operated continuously without observing another galactic event~\cite{Super-Kamiokande:2007zsl, KamLAND:2022sqb, Super-Kamiokande:2022dsn, IceCube:2023ogt}. The next nearby SN will provide unprecedented statistics in current and future detectors such as Super-Kamiokande~\cite{Super-Kamiokande:2007zsl}, DUNE~\cite{DUNE:2020txw}, JUNO~\cite{JUNO:2015zny}, and hyper-Kamiokande~\cite{Hyper-Kamiokande:2018ofw}, enabling precision tests of neutrino emission, flavor conversion, and fundamental neutrino properties~\cite{Janka:2017vlw, Scholberg:2012id}. In addition, multi-ton-scale dark matter detectors will reach comparable sensitivity~\cite{Lang:2016zhv}.

Neutrinos from SNe provide a unique laboratory for physics beyond the Standard Model. Since the origin of neutrino masses and mixings requires new interactions, nonstandard neutrino interactions (NSI) with matter have been widely studied, especially for vector and axial-vector couplings~\cite{Gavela:2008ra,  Antusch:2008tz, Ohlsson:2012kf, Proceedings:2019qno, Esteban:2018ppq} and for their impact on scattering rates~\cite{AristizabalSierra:2018eqm, Farzan:2018gtr, Giunti:2019xpr, Suliga:2020jfa}. In this work, we focus on scalar-mediated nonstandard interactions (SNSI), which naturally arise in extended Higgs sectors~\cite{Dutta:2020scq, Dutta:2022fdt, Dutta:2024hqq} and modify the neutrino mass matrix directly~\cite{Ge:2018uhz, Smirnov:2019cae, Babu:2019iml}. This behavior is fundamentally different from vector NSI, which modifies only the matter potential~\cite{Esteban-Pretel:2007zkv, Huang:2021enl, Denton:2022pxt, Jana:2024lfm}. In the Earth and Sun, the leading SNSI effects scale linearly with the ambient matter density~\cite{Denton:2024upc, Gupta:2023wct}. In contrast, we demonstrate that in SNe, the dominant SNSI contribution scales quadratically with density, leading to qualitatively new flavor evolution phenomena in the extreme SN environment.

The impact of SNSI in SNe has previously been explored in the context of collective oscillations~\cite{Blennow:2008er, Stapleford:2016jgz, Yang:2018yvk}, as well as through constraints based on potentially large medium-induced neutrino masses in the SN core~\cite{Smirnov:2019cae, Babu:2019iml}. Large induced masses could strongly affect neutrino production and the emitted spectra. In this work, we present the first study of SNSI effects on neutrino flavor evolution during the neutronization burst phase of a future galactic SN. This early-time $\nu_e$ flash has robust temporal and spectral features and is comparatively insensitive to astrophysical uncertainties that can strongly affect the flavor composition at Earth~\cite{Capanema:2024hdm}. A key signature is the anomalous preservation of the neutronization-burst $\nu_e$ peak in the normal mass ordering (NMO), in contrast to the standard Mikheyev-Smirnov-Wolfenstein (MSW)~\cite{Wolfenstein:1977ue, Mikheyev:1985zog, Mikheev:1986wj, Mikheev:1987jp} expectation of a suppressed peak. Using realistic SN profiles and the expected capabilities of DUNE~\cite{DUNE:2020zfm}, we show that flavor conversion during the neutronization burst can probe SNSI couplings 2 to 4 orders of magnitude below existing terrestrial and solar bounds, while accessing parameter regions not covered by studies relying on large induced masses.

%%%%%%%%%%%%%%%%%%%%%%%%%%%%%%%%%%%%%%%%%%%
%%%%%%%%%%%%%%%%%%%%%%%%%%%%%%%%%%%%%%%%%%%

\section{Neutronization Burst}

The neutronization burst is the earliest and most sharply defined phase of neutrino emission from a core-collapse SN~\cite{Mirizzi:2015eza, Janka:2017vlw}. It is driven by electron capture on free protons behind the shock front, producing an intense and short-lived $\nu_e$ flash within the first $\sim 50$~ms after core bounce, with typical energies of $10$--$20$~MeV and a luminosity exceeding that of the other flavors by 1 to 2 orders of magnitude. Owing to its short duration, high $\nu_e$ purity, and weak dependence on the progenitor and equation of state, this burst provides a particularly clean probe of flavor conversion and new physics~\cite{Dighe:1999bi, Kachelriess:2004ds, Serpico:2011ir, Wallace:2015xma, OConnor:2018sti, deGouvea:2019goq, Jana:2022tsa, Jana:2023ufy}. Throughout this work we assume a fiducial supernova-Earth distance of $R=10$ kpc. The time evolution and integrated spectra for the $12\,M_\odot$ progenitor adopted in this work~\cite{Garching_rep}, as well as the benchmark density profile at $\sim 5$~ms postbounce~\cite{Fischer:2009af, Tang:2020pkp}, are shown in the Supplemental Material~\cite{SupplementalMaterial}. These sharp temporal and spectral features make the burst especially well suited for detection in $\nu_e$-sensitive experiments such as DUNE~\cite{DUNE:2020zfm, Cuesta:2023nnt, Zhu:2018rwc}.

%%%%%%%%%%%%%%%%%%%%%%%%%%%%%%%%%%%%%%%%%%%
%%%%%%%%%%%%%%%%%%%%%%%%%%%%%%%%%%%%%%%%%%%
\section{Scalar nonstandard interactions}

SNSI arise when a new scalar mediator $\phi$ couples to neutrinos and matter fermions, modifying neutrino propagation in dense media~\cite{Ge:2018uhz, Smirnov:2019cae, Babu:2019iml}. The relevant Lagrangian is
\begin{align}
\mathcal{L} \supset & -\frac{1}{2}(\partial_\mu \phi)^2 - \frac{1}{2} m_\phi^2 \phi^2
+ \bar{\nu}_\alpha i\slashed{\partial}\nu_\beta
- m_{\alpha\beta}\bar{\nu}_\alpha\nu_\beta  \notag \\
& - y_{\alpha\beta}\bar{\nu}_\alpha\nu_\beta\phi
- y_f \bar{f} f \phi ,
\end{align}
where $m_\phi$ is the scalar mass, and $y_{\alpha\beta}$ and $y_f$ denote the Yukawa couplings to neutrinos and matter fermions.

In the contact-interaction limit assumed here, the scalar can be integrated out, yielding the effective interaction
\begin{equation}
\mathcal{L}^{\rm eff} \supset -\frac{y_f y_{\alpha\beta}}{m_\phi^2}
(\bar{\nu}_\alpha\nu_\beta)(\bar{f}f).
\end{equation}
This induces a medium-dependent correction to the neutrino mass matrix,
\begin{equation}
\delta M_{\alpha\beta}
= \sum_f \frac{N_f y_f y_{\alpha\beta}}{m_\phi^2},
\end{equation}
where $N_f\simeq \langle \bar f f\rangle$ is the fermion number density. Restricting to universal couplings to $u$ and $d$ quarks with strength $y_q$, one obtains
\begin{equation}
\delta M_{\alpha\beta}
= \frac{3 y_q y_{\alpha\beta}}{m_\phi^2}(N_p+N_n)
= \sqrt{\Delta m^2_{21}}\,\eta_{\alpha\beta}^\odot
\frac{\rho}{\rho_\odot}.
\end{equation}
Here $\rho_\odot = 100~{\rm g/cm}^3$ is the characteristic solar density scale. Couplings to electrons or other matter fermions would lead to analogous qualitative behavior, with the scalar source term rescaled by the corresponding matter density and coupling. For electron couplings, this source is reduced relative to the quark benchmark by \(N_e/[3(N_p+N_n)]=Y_e/3\), for the same assumed coupling strength\footnote{Possible scalar neutrino self-interactions~\cite{Yang:2018yvk} constitute an additional effect not included in the Hamiltonian above. Since the neutronization burst is initially dominated by \(\nu_e\), a flavor-nonuniversal \(\tau\tau\) feedback develops only after the additional conversion has generated a \(\nu_\tau\) component. As discussed in the Supplemental Material~\cite{SupplementalMaterial}, such feedback can shift the conversion to larger radii, but does not generically spoil the adiabatic flavor-conversion signature studied in this work.}. The dimensionless solar-normalized SNSI parameter is defined as\footnote{The superscript $\odot$ indicates normalization to the solar density scale and solar mass-squared differences~\cite{Denton:2024upc}. This definition may differ from those adopted in other works in the literature.}~\cite{Denton:2024upc}
\begin{equation} \label{eq:eq5}
\eta_{\alpha\beta}^\odot
= \frac{3 y_q y_{\alpha\beta} \rho_\odot}
{\sqrt{\Delta m^2_{21}} m_\phi^2 m_N}.
\end{equation}
Numerically,
\begin{equation}
\frac{y_{\alpha\beta} y_q}{m_\phi^2}
= 6.3 \times 10^{-15}\,{\rm eV}^{-2}\,\eta_{\alpha\beta}^\odot .
\end{equation}

The flavor-basis evolution Hamiltonian including SNSI is, therefore,
\begin{multline}
\mathcal{H}_F
= \frac{1}{2E}
\Big(
[U\,{\rm diag}(m_1,m_2,m_3)\,U^\dagger + \delta M]
\\[4pt]
\times
[U\,{\rm diag}(m_1,m_2,m_3)\,U^\dagger + \delta M]^\dagger
\Big)
+ {\rm diag}(V,0,0).
\label{eq:Hfull}
\end{multline}
where $\delta M$ is defined above. Unlike vector NSI, which modify only the matter potential, SNSI directly alter the neutrino mass matrix and introduce a dependence on the absolute neutrino mass scale. In the analysis below, we activate one $\eta_{\alpha\beta}^\odot$ parameter at a time to isolate its impact on SN flavor evolution.

%For comparison, vector-mediated NSI introduce an additive correction $V\to V+V_{\rm NSI}$ to the matter potential and have been extensively studied in the SN context~\cite{Esteban-Pretel:2007zkv, Jana:2024lfm}.
%%%%%%%%%%%%%%%%%%%%%%%%%%%%%%%%%%%%%%%%%%%
%%%%%%%%%%%%%%%%%%%%%%%%%%%%%%%%%%%%%%%%%%%

%%%%%%%%%%%%%%%%%%%%%%%%%%%%%%%%%%%%%%%%%%%
%%%%%%%%%%%%%%%%%%%%%%%%%%%%%%%%%%%%%%%%%%%
\section{Impact of a nonzero $\eta_{\tau\tau}^\odot$}

We begin by activating a single parameter and consider the case where only $\eta_{\tau\tau}^\odot$ (or $\delta M_{\tau\tau}$) is nonzero. The full Hamiltonian becomes
\begin{multline}
\mathcal{H}_F
= \frac{1}{2E}
U
\begin{pmatrix}
0 & 0 & 0 \\
0 & \Delta m_{21}^2 & 0 \\
0 & 0 & \Delta m_{31}^2
\end{pmatrix}
U^\dagger \\[4pt]
+ \frac{1}{2E}
\begin{pmatrix}
2EV & 0 & c_{e\tau}\delta M_{\tau\tau} \\
0 & 0 & c_{\mu\tau}\delta M_{\tau\tau} \\
c_{\tau e}\delta M_{\tau\tau} & c_{\tau\mu}\delta M_{\tau\tau}
& \delta M_{\tau\tau}^2 + 2 c_{\tau\tau}\delta M_{\tau\tau}
\end{pmatrix},
\label{eq:complete-H}
\end{multline}
where $c_{\alpha\beta}=\sum_j m_j\,U_{\alpha j}U_{\beta j}^*$ are the elements of the vacuum neutrino mass matrix in the flavor basis. Unlike standard matter effects, which depend linearly on the density $\rho$, the SNSI terms introduce both a dependence on the absolute neutrino mass scale through $c_{\alpha\beta}$ and quadratic-density contributions through $\delta M_{\tau\tau}^2$. Depending on the strength of $\eta_{\tau\tau}^\odot$, these terms can rearrange the level ordering, modify the MSW structure, or induce new resonances at small radii. An analogous evolution is obtained for $\eta_{\mu\mu}^\odot$, as discussed later.

\section{Different regimes with a nonzero $\eta_{\tau \tau}^\odot$}

The evolution governed by $\mathcal{H}_F$ exhibits distinct regimes controlled by the local matter density and the SNSI parameter $\eta_{\tau\tau}^\odot$. At densities $\rho \gtrsim 10^4~{\rm g/cm^3}$ ($r \lesssim 10^4~{\rm km}$), vacuum terms are negligible and will be omitted in the following estimates. Since both the standard matter potential and the linear SNSI terms proportional to $c_{\alpha\beta}\delta M_{\tau\tau}$ scale linearly with $\rho$, their ratio is density independent. For instance,
\begin{equation}
  \frac{2 c_{\tau \tau}  \delta M_{\tau \tau}}{2EV}
  \approx 6 \, \eta_{\tau \tau}^\odot ,
  \label{eq:ratio-1}
\end{equation}
where we adopt $E=10~{\rm MeV}$, $\Delta m^2_{21}=7.5\times10^{-5}~{\rm eV}^2$~\cite{Esteban:2024eli}, and $m_1<0.01~{\rm eV}$, yielding $c_{\tau\tau}\approx0.025~{\rm eV}$, $0.001~{\rm eV}\lesssim c_{e\tau}\lesssim0.008~{\rm eV}$ (varying $\delta_{\rm CP}$), and $c_{\mu\tau}\approx0.025~{\rm eV}$. The standard matter potential therefore dominates over linear SNSI terms as long as $|\eta_{\tau\tau}^\odot| \lesssim 0.1$.

The assumption $m_1<0.01~{\rm eV}$ is consistent with the cosmological bound $\sum m_\nu<0.072~{\rm eV}$~\cite{DESI:2024mwx} for NMO. With this assumption, the coefficients $c_{\alpha\beta}$ depend only weakly on the absolute neutrino mass scale. This constraint is substantially stronger than the kinematic bound from KATRIN, $m_1<0.45~{\rm eV}$~\cite{KATRIN:2024cdt}. The implications of relaxing this assumption are discussed in the Supplemental Material~\cite{SupplementalMaterial}.

The ratio between the standard matter potential and the quadratic SNSI term is instead
\begin{equation}
    \frac{2EV}{\delta M_{\tau \tau}^2}
    \approx \frac{1}{\eta_{\tau \tau}^{\odot 2} (\rho/\rho_\odot)} ,
    \label{eq:ratio-2}
\end{equation}
which varies with density/radius. Since neutrinos are emitted at $\rho \sim 10^{11}~{\rm g/cm^3}$ ($r\sim50$--$100$ km), for $|\eta_{\tau \tau}^\odot|\gtrsim10^{-4}$ the quadratic term initially dominates over $2EV$ and later becomes subdominant, potentially inducing an additional inner resonance between $\nu_e$ and $\nu_\tau$ prior to the standard MSW transitions. This will be discussed in the following sections.

By contrast, for $|\eta_{\tau\tau}^\odot|\gtrsim0.1$, the linear SNSI term $2c_{\tau\tau}\delta M_{\tau\tau}$ becomes comparable to, or larger than, $2EV$, modifying the MSW structure, with a strong dependence on the sign of $\eta_{\tau\tau}^\odot$ (this is explained in the Supplemental Material~\cite{SupplementalMaterial}). Finally, for $|\eta_{\tau\tau}^\odot|\ll10^{-4}$, all SNSI terms are negligible and $\mathcal{H}_F$ reduces to the standard oscillation in matter. 

For clarity, a visual comparison of the radial behavior of $2EV$, $2c_{\tau\tau}\delta M_{\tau\tau}$, and $\delta M_{\tau\tau}^2$, and of how they determine the different flavor-conversion regimes, is provided in the Supplemental Material~\cite{SupplementalMaterial}.

\section{Additional $\nu_e \leftrightarrow \nu_\tau$ resonant conversion induced by SNSI}

An additional resonance appears for $10^{-4} \lesssim |\eta_{\tau\tau}^\odot| < 0.1$. In this regime, the new inner resonance involves the $\nu_e$ and $\nu_\tau$ states, while the $\nu_\mu$ state acts essentially as a spectator, so the problem can be reduced to the following effective $\nu_e$--$\nu_\tau$ $2\times2$ Hamiltonian:
\begin{equation}
\mathcal{H}_F^{2\times 2} \approx
\frac{1}{2E}
\begin{pmatrix}
2EV & c_{e\tau} \delta M_{\tau\tau} \\
c_{\tau e} \delta M_{\tau\tau} & \delta M_{\tau\tau}^2
\end{pmatrix}.
\end{equation}
Since $\delta M_{\tau\tau}^2$ initially exceeds $2EV$ and decreases more rapidly with radius, a resonance occurs when $2EV=\delta M_{\tau\tau}^2$. This occurs at $\rho \gtrsim 10^4~{\rm g/cm^3}$ ($r \lesssim 10^4$~km), prior to the standard MSW layers.

The resonance density is
\begin{equation}
\rho_{\rm res} \approx
\frac{\rho_\odot^2}{\Delta m_{21}^2 \eta_{\tau\tau}^{\odot 2}}
\left(
\frac{2\sqrt{2} E G_F Y_e}{m_N}
\right).
\label{eq:res-cond}
\end{equation}
Efficient flavor conversion requires the resonance to be adiabatic, with the adiabaticity condition given by
\begin{equation}
\gamma =
\left|
\frac{4|\mathcal{H}_{e\tau}|^2}
{\dot{\mathcal{H}}_{\tau\tau}-\dot{\mathcal{H}}_{ee}}
\right|_{\rm res}
\approx
\frac{2|c_{e\tau}|^2}{E}
\left|
\frac{1}{\rho}\frac{d\rho}{dr}
\right|_{\rm res}^{-1}
> 1 ,
\label{eq:gamma}
\end{equation}
which depends sensitively on $\delta_{\rm CP}$ through $c_{e\tau}$.

To make the dependence of $\gamma$ on $\delta_{\rm CP}$ explicit, we write
\begin{equation}
|c_{e\tau}|^2
= c_{13}^2 \left(A^2 + B^2 - 2AB\cos\delta_{\rm CP}\right),
\label{eq:ce-tau}
\end{equation}
with
$A \equiv c_{23}s_{13}\left(c_{12}^2 m_1 + s_{12}^2 m_2 - m_3\right)$ and
$B \equiv c_{12}s_{12}s_{23}(m_1 - m_2)$.
For $m_1=0$, NMO, and $E=10$~MeV, both $A$ and $B$ are negative, so that $|c_{e\tau}|$ is maximized for $\delta_{\rm CP}=\pi$ and minimized near $\delta_{\rm CP}=0$.

Approximating the neutronization-burst density profile as $\rho\propto r^{-3}$, one finds
\begin{equation}
\left|
\frac{1}{\rho}\frac{d\rho}{dr}
\right|_{\rm res}^{-1}
=\frac{r_{\rm res}}{3},
\end{equation}
with $r_{\rm res}\approx6.5\times10^4\,\eta_{\tau\tau}^{2/3}$~km. Since $\gamma\propto |c_{e\tau}|^2 r_{\rm res}$, adiabaticity increases both with $\eta_{\tau\tau}^\odot$ and as $\delta_{\rm CP}\to\pi$.

\begin{figure}
\centering
\includegraphics[width=1\linewidth]{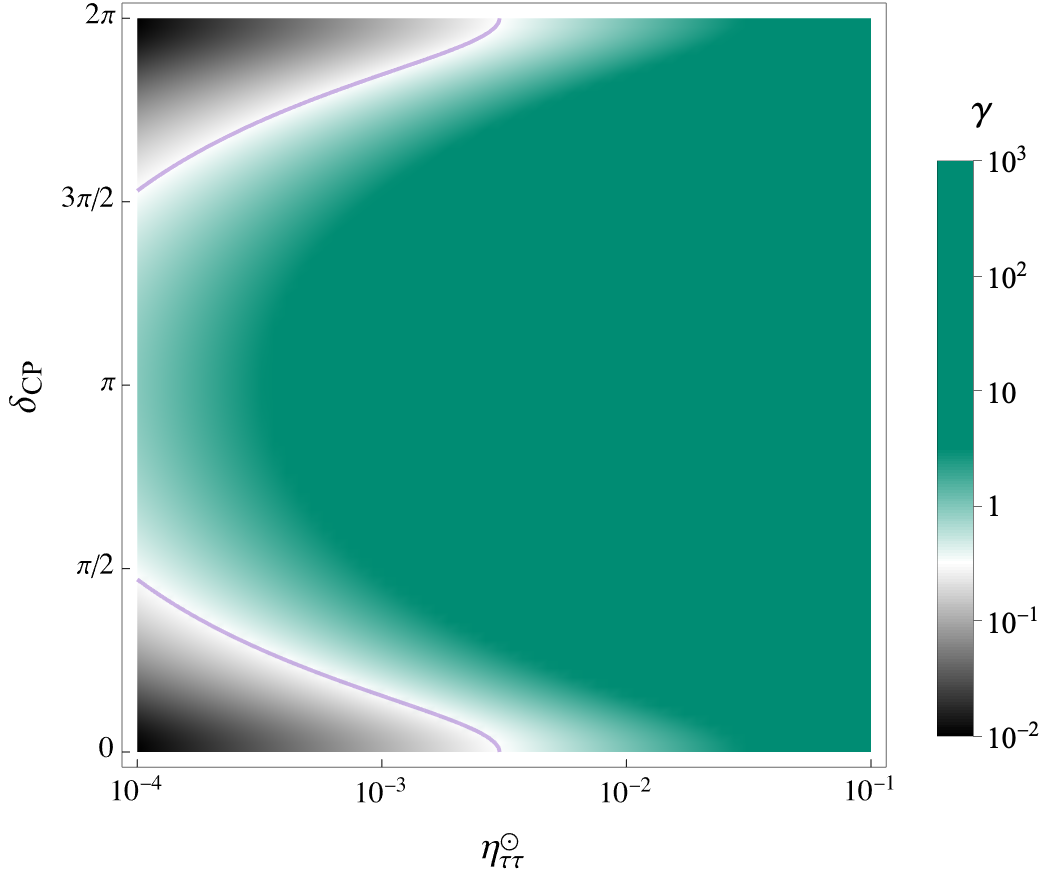}
\caption{Adiabaticity parameter $\gamma$ as a function of $\eta_{\tau\tau}^\odot$ and $\delta_{\rm CP}$, assuming $m_1=0$ and $E=10~{\rm MeV}$. Black regions correspond to fully nonadiabatic propagation ($\gamma\ll1$), green regions to fully adiabatic propagation ($\gamma\gg1$), and the color gradient denotes the transition regime. The purple contour marks the half-adiabatic condition $\gamma\simeq0.45$.}
\label{fig:gama-plot}
\end{figure}

The resulting adiabatic, nonadiabatic, and transition regimes are summarized in Fig.~\ref{fig:gama-plot}. The corresponding impact on detector event rates is shown in Fig.~\ref{fig:NO positive}.

%%%%%%%%%%%%%%%%%%%%%%%%%%%%%%%%%%%%%%%%%%%
%%%%%%%%%%%%%%%%%%%%%%%%%%%%%%%%%%%%%%%%%%%

%%%%%%%%%%%%%%%%%%%%%%%%%%%%%%%%%%%%%%%%%%%
%%%%%%%%%%%%%%%%%%%%%%%%%%%%%%%%%%%%%%%%%%%
\section{SNSI effects on DUNE event rates}

We now assess the observable consequences of a nonzero $\eta_{\tau\tau}^\odot$ on the $\nu_e$ flux at Earth and the corresponding DUNE event rates. For $|\eta_{\tau\tau}^\odot|<10^{-4}$, SNSI effects are negligible and the neutronization-burst $\nu_e$ is produced as the heaviest matter eigenstate, $\nu_3^m$, which propagates adiabatically and arrives at Earth as $\nu_3$. Since $\nu_3$ contains only a small $\nu_e$ component, the $\nu_e$ flux at a terrestrial detector is
\begin{equation}
\Phi_{\nu_e}^{\text{std}}
= |U_{e3}|^2 \Phi_{\nu_e}^i
+ \left(|U_{e1}|^2 + |U_{e2}|^2\right)\Phi_{\nu_x}^i
\approx 0.02\,\Phi_{\nu_e}^i + 0.98\,\Phi_{\nu_x}^i ,
\label{std-flux}
\end{equation}
where $\Phi_{\nu_x}^i$ denotes the common unoscillated $\nu_\mu$ and $\nu_\tau$ flux. Although $\Phi_{\nu_e}^i \approx 10\,\Phi_{\nu_x}^i$ at the neutronization peak, the $\nu_e$ contribution remains subdominant, and the peak is therefore suppressed for NMO. This standard case is shown by the black curve in Fig.~\ref{fig:NO positive} (details of the DUNE event-rate calculation and detector modeling are given in the Supplemental Material~\cite{SupplementalMaterial}).

For $|\eta_{\tau\tau}^\odot|\gtrsim10^{-4}$, the SNSI contribution to the $\nu_\tau$ potential dominates near the neutrinosphere, and $\nu_e$ is instead produced as the second-heaviest eigenstate, $\nu_2^m$. If the propagation is adiabatic, the initial $\nu_e$ flux arrives at Earth as $\nu_2$, yielding
\begin{equation}
\Phi_{\nu_e}^{\rm SNSI}
= |U_{e2}|^2 \Phi_{\nu_e}^i
+ (|U_{e1}|^2+|U_{e3}|^2)\Phi_{\nu_x}^i
\approx 0.3\,\Phi_{\nu_e}^i + 0.7\,\Phi_{\nu_x}^i .
\label{SNSI-flux}
\end{equation}
Since during the neutronization burst $\Phi_{\nu_e}^i \gg \Phi_{\nu_x}^i$, the peak remains partially visible. This behavior is illustrated by the green curves in Fig.~\ref{fig:NO positive}. For best-fit oscillation parameters~\cite{Esteban:2024eli}, the propagation is fully adiabatic for all $\eta_{\tau\tau}^\odot$ shown.

If the SNSI-induced resonance is not fully adiabatic, the initially produced $\nu_2^m$ can partially convert to $\nu_3^m$, and the final flux interpolates between the SNSI and standard cases,
\begin{equation}
\Phi_{\nu_e}
= (1-P_f)\Phi_{\nu_e}^{\rm SNSI}
+ P_f\,\Phi_{\nu_e}^{\rm std},
\label{eq:general-flux}
\end{equation}
with the Landau-Zener flip probability
$P_f = e^{-\frac{\pi}{2}\gamma}$~\cite{Landau:1932vnv, Zener:1932ws, Parke:1986jy}.
This intermediate regime is illustrated by the purple curve in Fig.~\ref{fig:NO positive}. In this case, the neutronization peak is partially suppressed but remains distinguishable. For $|\eta^\odot_{\tau\tau}|\gtrsim 0.1$, Eq.~\eqref{eq:general-flux} is effectively realized with $P_f=0$; see the Supplemental Material~\cite{SupplementalMaterial}.

Theoretical uncertainties in the \(\nu_e\)--\(^{40}\)Ar cross section~\cite{DUNE:2023rtr} and in the overall supernova flux normalization may affect the absolute normalization of the DUNE event rates, but they do not alter the main signature studied here, namely the presence or absence of the neutronization-burst peak in time. To quantify this point, we perform a shape-only comparison of the time distributions in the left panel of Fig.~\ref{fig:NO positive}, fully marginalizing over the overall normalization; details are given in the Supplemental Material~\cite{SupplementalMaterial}. Relative to the standard MSW prediction, we find \(\chi^2\simeq 38\) for the fully adiabatic SNSI case and \(\chi^2\simeq 17\) for the partially adiabatic case. The dominant contribution comes from the neutronization-burst peak around \(5\)-\(10\,{\rm ms}\), showing that the SNSI-induced restoration of the peak remains distinguishable even without relying on the absolute event normalization.

\begin{figure*}
   
    \includegraphics[width=0.49\linewidth]{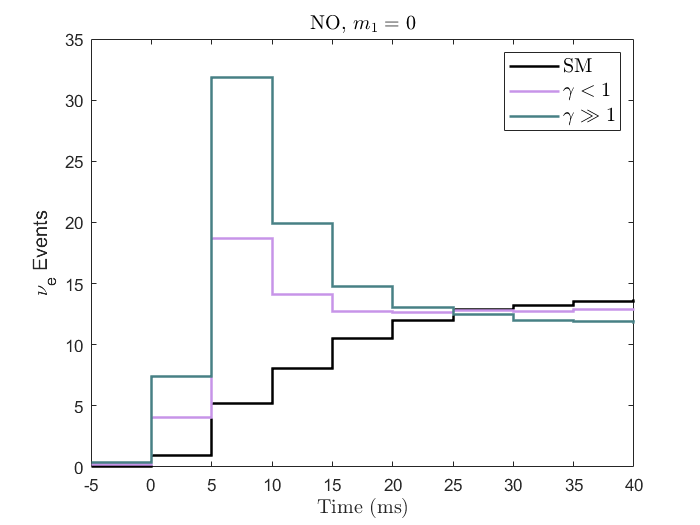}
    \includegraphics[width=0.49\linewidth]{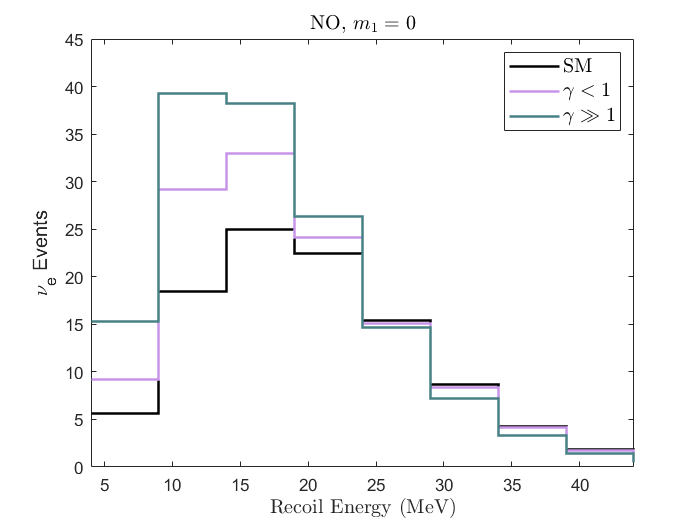}
    \caption{Expected $\nu_e$ event rates in DUNE for a $12\,M_\odot$ progenitor during the neutronization burst for NMO and $m_1=0$. The black curve corresponds to the standard MSW scenario (no SNSI or $\gamma \ll 1$). The green curve shows the fully adiabatic SNSI regime ($\gamma \gg 1$), where the neutronization peak is partially preserved, while the purple curve corresponds to the partially adiabatic SNSI regime ($\gamma \lesssim 1$), in which the peak is partially washed out. The three cases correspond to the adiabaticity regimes identified in Fig.~1. The left (right) panel shows the time (recoil-energy) distribution of events.}
    \label{fig:NO positive}
\end{figure*}
%%%%%%%%%%%%%%%%%%%%%%%%%%%%%%%%%%%%%%%%%%%
%%%%%%%%%%%%%%%%%%%%%%%%%%%%%%%%%%%%%%%%%%%

\section{Parameter region and existing constraints}

\begin{figure}[!t]
    \centering
    \includegraphics[width=\columnwidth]{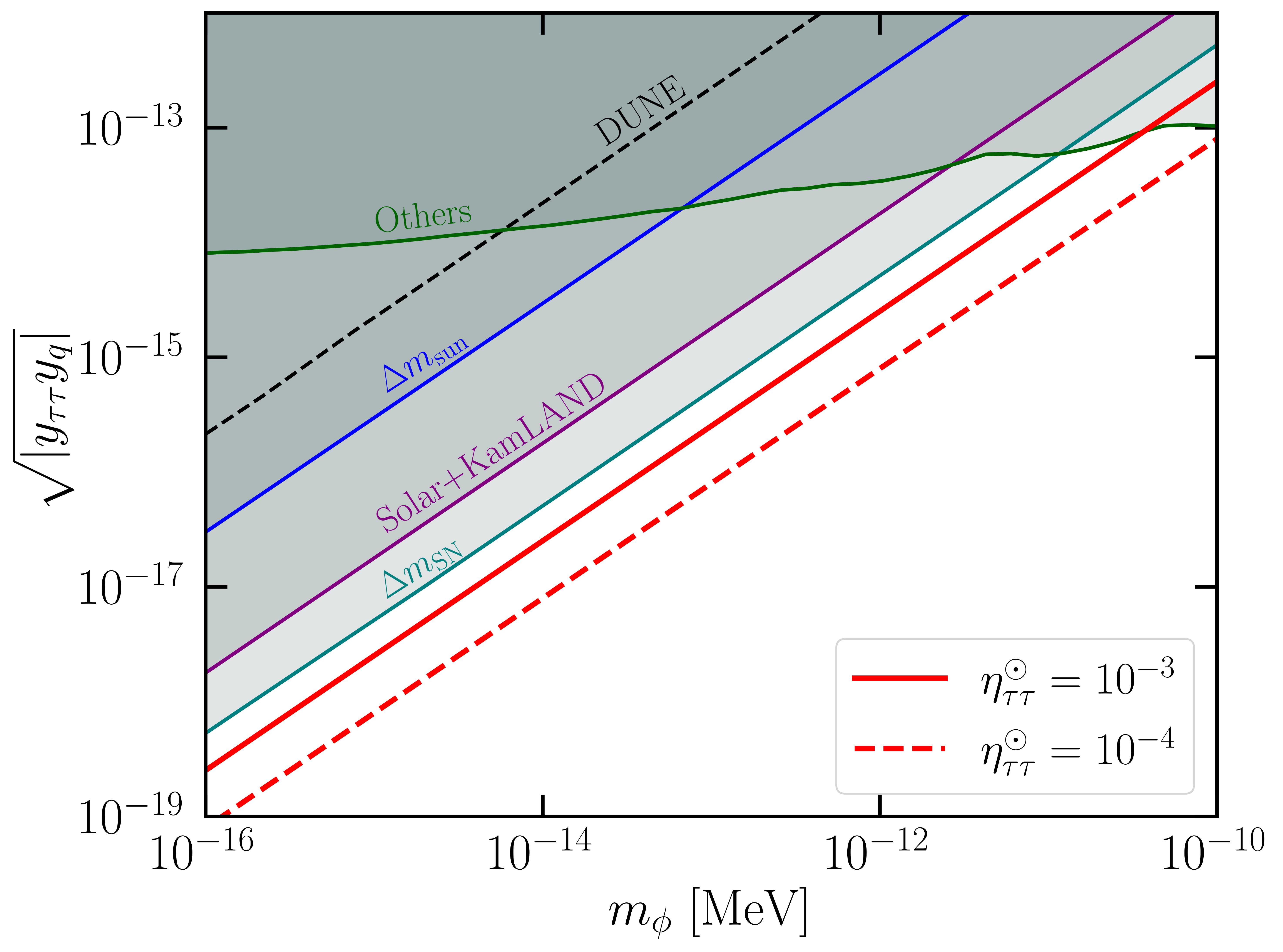}
    \caption{Parameter region in which the additional SNSI-induced flavor conversion discussed in this work occurs, shown in the scalar mass--coupling plane \((m_\phi,\sqrt{y_{\tau\tau}y_q})\) using the relation in Eq.~\eqref{eq:eq5}, with \(y_u=y_d=y_q\). The two dashed lines correspond to \(|\eta_{\tau\tau}^{\odot}|=10^{-3}\) and \(10^{-4}\). The bounds \(\Delta m_{\text{sun}}\) and \(\Delta m_{\text{SN}}\) depict constraints from in-medium neutrino masses inside the Sun and SN1987A, respectively~\cite{Babu:2019iml}. We also show constraints from BBN, \(\Delta N_{\text{eff}}\), and fifth-force experiments, collectively labeled as ``others''~\cite{Dutta:2024hqq}, as well as the current \(95\%\) C.L. bound on \(\eta_{\tau\tau}^{\odot}\) from combined solar and KamLAND data~\cite{Denton:2024upc}. Additionally, we show the projected \(95\%\) C.L. sensitivity of the upcoming long-baseline DUNE experiment~\cite{Dutta:2024hqq}. The region with \(m_\phi > 10^{-10}\,\mathrm{MeV}\) and \(\sqrt{y_{\tau\tau}y_q}>10^{-12}\) is omitted for presentation purposes only, since the relevant bounds there are strongly model dependent and this does not reflect a limitation of the validity of our analysis.}
    \label{fig:bounds_tautau}
\end{figure}

Fig.~\ref{fig:bounds_tautau} shows the parameter region in which the additional SNSI-induced flavor conversion discussed in this work occurs, projected through Eq.~\eqref{eq:eq5} onto the parameter space of scalar mass and coupling \((m_\phi,\sqrt{y_{\tau\tau}y_q})\), alongside existing constraints and projections from other searches. Here we use \(y_u=y_d=y_q\). In the same parameter space, we depict the constraints on the in-medium mass of neutrinos in the Sun and SN1987A~\cite{Babu:2019iml}, which directly probes $\sqrt{y_{\tau\tau}y_q}$ as a function of the scalar mass. The bounds labeled as ``others'' depict bounds from fifth-force experiments~\cite{Wise:2018rnb} as well as other cosmological bounds. We also depict the most stringent bound on \(|\eta_{\tau\tau}^{\odot}|=0.06\) from combined solar and KamLAND data~\cite{Denton:2024upc}. Alongside, we also show the projected sensitivity of the long-baseline DUNE experiment~\cite{Dutta:2024hqq}. 
We restrict our analysis to scalar masses $m_\phi > 10^{-16}~\text{MeV}$. Scalars lighter than this would mediate forces with macroscopic ranges ($\gtrsim 10$ km) comparable to the size of the SN core, leading to additional long-range contributions to the matter potential that require a dedicated treatment.
We also refrain from including the parameter space for $m_{\phi} > 10^{-10}~\text{MeV}$ and $\sqrt{y_{\tau\tau}y_q} > 10^{-12}$ where nontrivial model-dependent bounds from BBN, SN1987A, and stellar cooling~\cite{Hardy:2016kme}, SN1987A trapping~\cite{Heurtier:2016otg}, and neutrino self interactions~\cite{Fiorillo:2022cdq, Fiorillo:2023cas, Fiorillo:2023frv, Fiorillo:2023ytr, Telalovic:2024cot} become relevant. 

%%%%%%%%%%%%%%%%%%%%%%%%%%%%%%%%%%%%%%%%%%%
%%%%%%%%%%%%%%%%%%%%%%%%%%%%%%%%%%%%%%%%%%%
\section{SN versus solar neutrinos in probing SNSI} Exploiting the neutronization burst, SN neutrinos can probe \(|\eta_{\tau\tau}^{\odot}|\) down to \(\sim 10^{-4}\), well below the values currently accessible to solar neutrinos, \(|\eta_{\tau\tau}^{\odot}|\sim O(1)\)~\cite{Denton:2024upc}. For solar neutrinos, rewriting Eq.~\eqref{eq:ratio-1} yields
\begin{equation}
  \frac{2 c_{\tau \tau} \, \delta M_{\tau \tau}}{2EV} \approx 12\,\eta_{\tau \tau}^\odot ,
  \label{eq:ratio-1-sun}
\end{equation}
where the numerical factor reflects a characteristic solar neutrino energy of $5$~MeV. As a result, SNSI effects become relevant only for $|\eta_{\tau\tau}^\odot|\gtrsim0.1$, consistent with existing solar bounds. In this regime, the quadratic term remains subdominant, as
\begin{equation}
  \frac{2 c_{\tau \tau}\,\delta M_{\tau \tau}}{\delta M_{\tau \tau}^2}
  \approx \frac{6}{\eta_{\tau\tau}^\odot(\rho/\rho_\odot)} \gg 1
  \label{eq:ratio-2-sun}
\end{equation}
at solar densities and grows further as $\rho$ decreases. Solar sensitivity is therefore controlled by the linear SNSI term.

By contrast, the SN reach is driven by the quadratic term $\delta M_{\tau\tau}^2$, strongly enhanced by the extreme densities near the neutrinosphere. This qualitative difference in density scaling underlies the orders-of-magnitude improvement of SN neutrinos over solar probes.
%%%%%%%%%%%%%%%%%%%%%%%%%%%%%%%%%%%%%%%%%%%
%%%%%%%%%%%%%%%%%%%%%%%%%%%%%%%%%%%%%%%%%%%

%%%%%%%%%%%%%%%%%%%%%%%%%%%%%%%%%%%%%%%%%%%
%%%%%%%%%%%%%%%%%%%%%%%%%%%%%%%%%%%%%%%%%%%
\section{Final remarks}

We have shown that neutrinos from a future galactic core-collapse SN provide an exceptional probe of SNSI. Owing to the quadratic density scaling of SNSI effects, the neutronization burst can exhibit an anomalous $\nu_e$ peak in the NMO, even for couplings well below the current reach of solar and terrestrial experiments. With its excellent time and energy resolution, DUNE can test values as small as \(|\eta_{\tau\tau}^{\odot}|\sim10^{-4}\), highlighting the next nearby SN as a unique natural laboratory for scalar-mediated new physics. For vector NSI, the density-squared contribution studied in this work is absent; an analogous burst-revival signature would instead require non-electron flavor diagonal NSI above \(O(0.1)\), together with off diagonal entries sufficient for adiabatic conversion~\cite{Jana:2024lfm}, parameter combinations constrained or directly tested by non-supernova oscillation and CE\(\nu\)NS data~\cite{Coloma:2023ixt}.

In the inverted mass ordering, both the Standard Model and SNSI predict the presence of a neutronization peak, leading to a strong degeneracy and a substantially reduced ability to distinguish SNSI from the standard expectation. As a result, the discriminating power of the neutronization burst is maximized in the NMO, where the appearance of the peak would constitute a clear signal of beyond-Standard-Model physics.

During the period of our publication, a complementary study of SNSI in SN neutrinos also appeared in Ref.~\cite{Das:2025zts}, with emphasis on detector event rates and the associated mass-ordering ambiguity.
%%%%%%%%%%%%%%%%%%%%%%%%%%%%%%%%%%%%%%%%%%%
%%%%%%%%%%%%%%%%%%%%%%%%%%%%%%%%%%%%%%%%%%%

\vspace{0.2in}
%%%%%%%%%%%%%%%%%%%%%%%%%%%%%%%%%%%%%%%%%%%
%\begin{acknowledgments}
\begin{acknowledgments}
We are grateful to Peter Denton, P.~S. Bhupal Dev, and Anna Suliga for helpful discussions and suggestions. The work of B.~D., A.~K., N.~M., and L.~S. is supported by the U.S.~DOE Grant No.~DE-SC0010813. The work of Y.~P. was supported by the São Paulo Research Foundation (FAPESP) under Grants No.~2023/10734-3 and No.~2023/01467-1, and by the National Council for Scientific and Technological Development (CNPq) under Grant No.~151168/2023-7. Y.~P. gratefully acknowledges the hospitality of Texas A\&M University, where this collaboration was initiated.
\end{acknowledgments}
 %%%%%%%%%%%%%%%%%%%%%%%%%%%%%%%%%%%%%%%%%%%

 \section*{Data availability}

The data that support the findings of this article are not publicly available because they are owned by a third party and the terms of use prevent public distribution. The data are available from the authors upon reasonable request.

%%%%%%%%%%%%%%%%%%%%%%%%%%%%%%%%%%%%%%%%%%%
\bibliographystyle{utcaps_mod}
\bibliography{refs}

\clearpage

% Independent supplemental numbering.
\setcounter{section}{0}
\setcounter{subsection}{0}
\setcounter{equation}{0}
\setcounter{figure}{0}
\setcounter{table}{0}
\renewcommand{\thesection}{S\arabic{section}}
\renewcommand{\thesubsection}{\thesection.\Alph{subsection}}
\renewcommand{\theequation}{S\arabic{equation}}
\renewcommand{\thefigure}{S\arabic{figure}}
\renewcommand{\thetable}{S\arabic{table}}

% Unique PDF hyperlink destinations.
\providecommand{\theHsection}{}
\providecommand{\theHsubsection}{}
\providecommand{\theHequation}{}
\providecommand{\theHfigure}{}
\providecommand{\theHtable}{}
\renewcommand{\theHsection}{supp.\arabic{section}}
\renewcommand{\theHsubsection}{supp.\arabic{section}.\arabic{subsection}}
\renewcommand{\theHequation}{supp.\arabic{equation}}
\renewcommand{\theHfigure}{supp.\arabic{figure}}
\renewcommand{\theHtable}{supp.\arabic{table}}

\onecolumngrid
\section*{Supplemental Material}
\label{sec:supplement}
\twocolumngrid

%%%%%%%%%%%%%%%%%%%%%%%%%%%%%%%%%%%%%%%%%%%%%%%%%%%%%%%%%%%%
\section{Neutronization-burst input model}
%%%%%%%%%%%%%%%%%%%%%%%%%%%%%%%%%%%%%%%%%%%%%%%%%%%%%%%%%%%%

\begin{figure*}[t]
  \centering
  \includegraphics[width=0.48\textwidth]{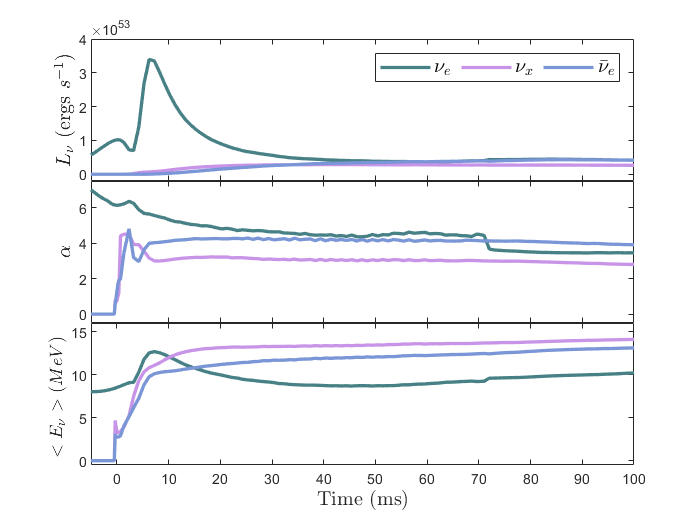}
  \hfill
  \includegraphics[width=0.48\textwidth]{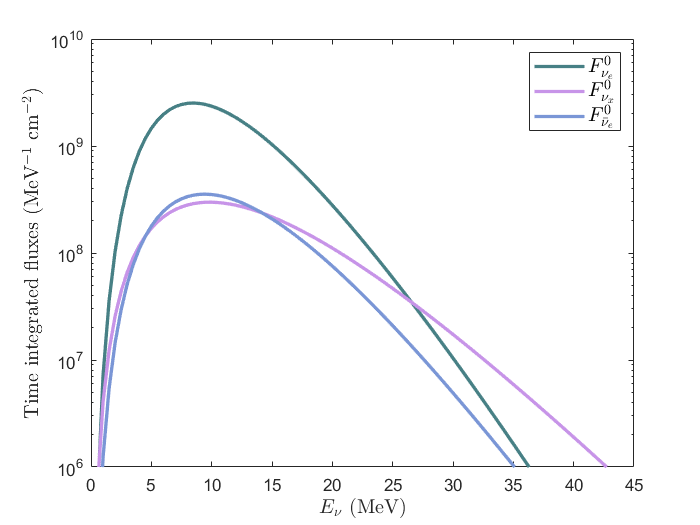}
  \caption{\textbf{Left}: Time evolution of the spectral parameters for the $12\,M_\odot$ Garching model used in this work~\cite{Garching_rep}.
  \textbf{Right}: Fluxes from Eq.~\eqref{eq:initial-fluxes-sm} integrated over the first $\sim 50~\text{ms}$, illustrating the dominance of $\nu_e$ during the burst.}
  \label{fig:flux_sm}
\end{figure*}

We model the neutronization-burst emission using the $12\,M_\odot$ Garching simulation data for the time-dependent
luminosities, average energies, and pinching parameters (Fig.~\ref{fig:flux_sm})~\cite{Garching_rep}. We assume a fiducial
supernova distance of $R=10~\text{kpc}$.

For each flavor $i=\nu_e,\bar\nu_e,\nu_x$, the unoscillated differential flux at Earth is parameterized as
\begin{equation}
\begin{aligned}
\Phi_{\nu}^i(E,t)={}&\frac{1}{4\pi R^2}\,
\frac{L_\nu(t)}{\langle E_\nu\rangle^2}\,
\frac{(\alpha+1)^{\alpha+1}}{\Gamma(\alpha+1)}
\\
&{}\times\left(\frac{E}{\langle E_\nu\rangle}\right)^{\alpha}
\exp\!\left[-(\alpha+1)\frac{E}{\langle E_\nu\rangle}\right],
\end{aligned}
\label{eq:initial-fluxes-sm}
\end{equation}
where $L_\nu(t)$, $\langle E_\nu\rangle(t)$, and $\alpha(t)$ are taken from the simulation output.

The benchmark matter profile used for flavor evolution is the $\sim 5~\text{ms}$ postbounce density profile adopted in the main text~\cite{Fischer:2009af}.

%%%%%%%%%%%%%%%%%%%%%%%%%%%%%%%%%%%%%%%%%%%%%%%%%%%%%%%%%%%%
\section{Modification of energy levels in the MSW layers for \texorpdfstring{$\left|\eta^\odot_{\tau\tau}\right|\gtrsim0.1$}{absolute eta greater than or approximately equal to 0.1}}
\label{sec:modify-MSW}
%%%%%%%%%%%%%%%%%%%%%%%%%%%%%%%%%%%%%%%%%%%%%%%%%%%%%%%%%%%%

In this regime, the effect of scalar nonstandard interactions (SNSI) is highly sensitive to the sign of $\eta_{\tau \tau}^\odot$.

\subsection{Positive SNSI (\texorpdfstring{$\eta^\odot_{\tau \tau} \gtrsim 0.1$}{eta greater than or approximately equal to 0.1})}

For $\eta^\odot_{\tau \tau} \gtrsim 0.1$, the resonance induced by SNSI does not occur, even though the condition
$2EV=\delta M_{\tau \tau}^2$ is still satisfied in the outer layers ($\rho \lesssim 10^4~\text{g/cm}^3$). The reason is that the
contribution from $c_{\tau \tau}\,\delta M_{\tau \tau}$ (as well as the vacuum terms) is no longer negligible and keeps the $\nu_\tau$
energy level higher than that of $\nu_e$ throughout, even after $\delta M_{\tau \tau}^2$ becomes much smaller than $2EV$.

To illustrate this, consider $\eta_{\tau \tau}^\odot > 0.2$ and, for didactic purposes, neglect the vacuum terms. The system is then
described by the approximate Hamiltonian
\begin{multline}
\mathcal{H}_F^{2\times2}\approx
\\
\frac{1}{2E}
\begin{pmatrix}
2EV & c_{e\tau}\delta M_{\tau\tau} \\
c_{\tau e}\delta M_{\tau\tau} & \delta M_{\tau\tau}^2+2c_{\tau\tau}\delta M_{\tau\tau}
\end{pmatrix}.
\label{eq:2x2-outer-layers}
\end{multline}
In this case, even when $\delta M_{\tau \tau}^2 \ll 2EV$, the energy levels of $\nu_e$ and $\nu_\tau$ do not cross: for sufficiently large
$\eta^\odot_{\tau\tau}$ the term $2c_{\tau \tau}\,\delta M_{\tau \tau}$ dominates over $2EV$ and keeps the $\nu_\tau$ level higher at all densities.
Including the vacuum terms, the onset of this behavior shifts to $\eta_{\tau \tau}^\odot \gtrsim 0.1$. Furthermore, since no
$\nu_e \leftrightarrow \nu_\tau$ transition occurs in the MSW layers, the standard $H$-resonance is absent.
Therefore, although SNSI do not induce a new resonance for $\eta^\odot_{\tau \tau} \gtrsim 0.1$, they still modify the energy levels in the outer
layers and alter the standard MSW dynamics.

\subsection{Negative SNSI (\texorpdfstring{$\eta^\odot_{\tau \tau} \lesssim -0.1$}{eta less than or approximately equal to -0.1})}

For $\eta_{\tau \tau}^\odot \lesssim -0.1$, the energy levels are also modified in the outer layers, but unlike the case of positive SNSI, a new
resonance can still be present. This occurs because the negative contribution from $2 c_{\tau \tau}\,\delta M_{\tau \tau}$ drives the $\nu_\tau$ energy
level below that of $\nu_e$ [see Eq.~\eqref{eq:2x2-outer-layers}] shortly after $\delta M_{\tau \tau}^2$ becomes smaller than $2EV$. This level crossing
persists even when vacuum terms are included. In this case, the SNSI-induced resonance may occur in proximity to the standard MSW resonances, making a
numerical treatment more reliable than a purely analytical one.

There is an additional caveat in the negative SNSI scenario: due to the negative contribution from $2c_{\tau \tau}\,\delta M_{\tau \tau}$, the $\nu_\tau$
energy level can drop not only below that of $\nu_e$, but also below the $\nu_\mu$ level. The resulting $\nu_\tau \leftrightarrow \nu_\mu$ level crossing
typically occurs rapidly and is generally not fully adiabatic. Although this crossing may not result in efficient flavor conversion, it can still induce fast
neutrino oscillations inside the supernova due to the nonadiabaticity, which projects the neutrino state onto a linear combination of distinct instantaneous
eigenstates after the crossing.

%%%%%%%%%%%%%%%%%%%%%%%%%%%%%%%%%%%%%%%%%%%%%%%%%%%%%%%%%%%%
\section{Relative size of the standard matter, linear SNSI, and quadratic SNSI terms}
%%%%%%%%%%%%%%%%%%%%%%%%%%%%%%%%%%%%%%%%%%%%%%%%%%%%%%%%%%%%

\begin{figure*}[t]
    \centering
    \includegraphics[width=0.48\textwidth]{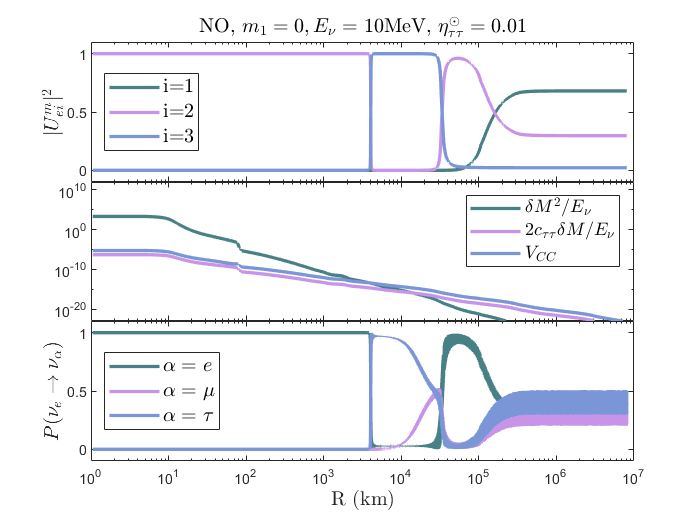}
    \hfill
    \includegraphics[width=0.48\textwidth]{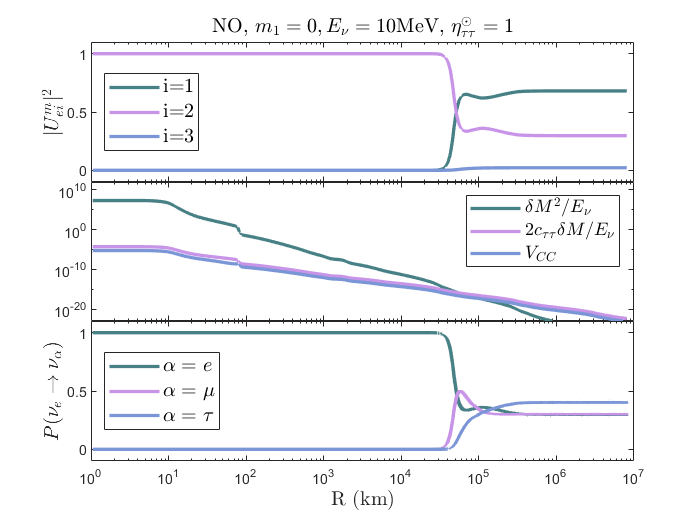}
    \caption{
    Impact of SNSI on flavor evolution in a supernova environment. \textbf{Left panels:} $\eta^\odot_{\tau\tau}=0.01$. \textbf{Right panels:} $\eta^\odot_{\tau\tau}=1$. In all panels we assume normal ordering, $m_1=0$, and $E_\nu=10$ MeV.
    \textbf{Top panels:} Squared moduli of the effective mixing matrix elements in matter, $|U^m_{ei}|^2$, as functions of radius $R$. For $\eta^\odot_{\tau\tau}=0.01$, an additional inner resonance appears while leaving the standard MSW region essentially unchanged. For $\eta^\odot_{\tau\tau}=1$, no additional inner resonance occurs, although the MSW configuration is substantially modified.
    \textbf{Middle panels:} Comparison of the quadratic SNSI term $\delta M_{\tau\tau}^2/E_\nu$, the linear SNSI contribution $2c_{\tau\tau}\delta M_{\tau\tau}/E_\nu$, and the standard matter term $V_{CC}$. For $\eta^\odot_{\tau\tau}=0.01$, the condition $\delta M_{\tau\tau}^2 = 2EV$ generates the new resonance. For $\eta^\odot_{\tau\tau}=1$, the same crossing still occurs, but no resonance is produced because the linear term keeps the $\nu_\tau$ level above the $\nu_e$ level.
    \textbf{Bottom panels:} Transition probabilities $P(\nu_e\rightarrow \nu_\alpha)$ as functions of radius. In the left panels, the disappearance of $\nu_e$ and the corresponding appearance of $\nu_\tau$ between $10^3$ and $10^4$ km signal the new resonance. In both cases, the $\nu_e$ survival probability is notably enhanced compared to the standard normal-ordering scenario shown in Fig.~\ref{fig:NO positive} of the main text.
    }
    \label{fig:terms_eta_tautau}
\end{figure*}

The different flavor-conversion regimes discussed in the main text can be understood by comparing the radial behavior of the three dominant contributions entering the effective $\nu_e$--$\nu_\tau$ system: the standard matter term $2EV$, the linear SNSI term $2c_{\tau\tau}\delta M_{\tau\tau}$, and the quadratic SNSI term $\delta M_{\tau\tau}^2$. This comparison is shown in Fig.~\ref{fig:terms_eta_tautau} for two representative values of $\eta^\odot_{\tau\tau}$.

The \textbf{top panels} show the squared moduli of the effective mixing matrix elements in matter, $|U^m_{ei}|^2$, as functions of radius. These identify which instantaneous matter eigenstate carries the electron flavor along the propagation. The \textbf{middle panels} show directly the three quantities that control the discussion in the main text, namely $\delta M_{\tau\tau}^2/E_\nu$, $2c_{\tau\tau}\delta M_{\tau\tau}/E_\nu$, and $V_{CC}$. Finally, the \textbf{bottom panels} display the corresponding transition probabilities $P(\nu_e\to \nu_\alpha)$, showing how the change in the level structure is reflected in the flavor content of the state.

The left panels correspond to $\eta^\odot_{\tau\tau}=0.01$, which lies in the range $10^{-4}\lesssim |\eta^\odot_{\tau\tau}|<0.1$. In this case, the quadratic SNSI term dominates near the neutrinosphere, but because it scales as $\rho^2$ it decreases more rapidly with radius than the standard matter term, which scales linearly with $\rho$. As seen in the middle-left panel, the condition $\delta M_{\tau\tau}^2=2EV$ is therefore satisfied at an inner radius. This produces an additional inner resonance, visible in the top-left panel as the exchange in the matter-eigenstate composition and in the bottom-left panel as the disappearance of $\nu_e$ together with the corresponding appearance of $\nu_\tau$ between $10^3$ and $10^4$ km. In this regime, the standard MSW structure at larger radii remains essentially unchanged.

The right panels correspond to $\eta^\odot_{\tau\tau}=1$, i.e.\ to the regime $\eta^\odot_{\tau\tau}\gtrsim 0.1$. In this case, although the condition $\delta M_{\tau\tau}^2=2EV$ is still satisfied, no new resonance appears. The reason is that the linear SNSI term $2c_{\tau\tau}\delta M_{\tau\tau}$ is already comparable to or larger than $2EV$, so that even after the quadratic term becomes subdominant, the $\nu_\tau$ level remains above the $\nu_e$ level and no level crossing is generated. This is reflected in the top-right panel, where no additional inner exchange of matter eigenstates occurs, and in the bottom-right panel, where the flavor evolution changes without exhibiting the extra resonance seen in the left panels. Thus, for large enough positive $\eta^\odot_{\tau\tau}$, SNSI modifies the standard MSW configuration but does not generate the additional inner resonance.

%%%%%%%%%%%%%%%%%%%%%%%%%%%%%%%%%%%%%%%%%%%%%%%%%%%%%%%%%%%%
\section{Relaxing the assumption \texorpdfstring{$m_1<0.01~\text{eV}$}{m1 less than 0.01 eV}}
%%%%%%%%%%%%%%%%%%%%%%%%%%%%%%%%%%%%%%%%%%%%%%%%%%%%%%%%%%%%

As $m_1$ increases into the quasi-degenerate regime, the three neutrino masses in vacuum become approximately equal,
i.e., $m_2 \approx m_3 \approx m_1$. At the same time, the values of $c_{\alpha \beta}$ become strongly dependent on the absolute mass scale $m_1$, following
$c_{\alpha \beta} \approx m_1 \sum_j U_{\alpha j} U_{\beta j}^*$. In particular, $c_{\tau \tau} \approx m_1 \sum_j |U_{\tau j}|^2 = m_1$, and previous
estimates can change significantly for $m_1 \gg 0.01$~eV. For instance,
\begin{equation}
  \frac{2 c_{\tau \tau}\,\delta M_{\tau \tau}}{2EV}
  \approx \left( \frac{m_1}{0.45~\text{eV}} \right) 100 \ \eta_{\tau \tau}^\odot ,
\label{eq:ratio-3}
\end{equation}
showing that, for $m_1 = 0.45$~eV, $2 c_{\tau \tau}\,\delta M_{\tau \tau}$ already dominates over $2EV$ when $\eta_{\tau \tau}^\odot \approx 10^{-2}$,
one order of magnitude smaller than in the $m_1=0$ case. In this regime, the evolution is qualitatively similar to the situation where SNSI modifies the energy
levels in the outer layers and can remove the new resonance. By contrast, for $|\eta_{\tau \tau}^\odot| < 10^{-2}$, $2EV$ dominates and a new SNSI-induced
resonance can emerge.

Another important feature of scenarios with larger $m_1$ is that the SNSI-induced resonances become less adiabatic as the mass increases. For instance, while
for $m_1 < 0.01$~eV we can have $c_{e\tau}$ as large as $0.008$~eV, for $m_1 = 0.45$~eV we find $c_{e\tau} \approx 0.0002$~eV. This suppresses the adiabaticity
relative to the case with $m_1 < 0.01$~eV. Therefore, for $m_1 \gtrsim 0.01~\text{eV}$, adiabaticity decreases progressively with increasing $m_1$, which suppresses
the sensitivity of supernova neutrino flavor evolution to SNSI that induce a new resonance, i.e., for $|\eta_{\tau \tau}^\odot| \ll 10^{-2}$.

%%%%%%%%%%%%%%%%%%%%%%%%%%%%%%%%%%%%%%%%%%%%%%%%%%%%%%%%%%%%
\section{Event-rate computation at DUNE}
%%%%%%%%%%%%%%%%%%%%%%%%%%%%%%%%%%%%%%%%%%%%%%%%%%%%%%%%%%%%

Electron neutrinos are detected at the DUNE Far Detector mainly via the charged-current process
$\nu_e+{}^{40}\text{Ar}\to e^-+{}^{40}\text{K}^\ast$.
We adopt a threshold $Q=1.5045~\text{MeV}$ (to the ground state), and define the electron kinetic energy as
$T_e(E_\nu)=E_\nu-Q-m_e$.
Detector resolution is included through a Gaussian smearing from the true neutrino energy to the visible energy,
\begin{equation}
W(E_{\rm vis},E_\nu)=\frac{1}{\Delta(E_{\rm vis})\sqrt{2\pi}}
\exp\!\left[
-\frac{(E_{\rm vis}-T_e(E_\nu))^2}{2\,\Delta(E_{\rm vis})^2}
\right],
\end{equation}
with
\begin{equation}
\frac{\Delta(E_{\rm vis})}{\text{MeV}}=
0.11\sqrt{\frac{E_{\rm vis}}{\text{MeV}}}
+0.02\frac{E_{\rm vis}}{\text{MeV}},
\end{equation}
consistent with the DUNE MeV-performance studies used in the main text~\cite{DUNE:2020zfm,Zhu:2018rwc}.

The differential number of events as a function of visible energy is
\begin{equation}
\begin{aligned}
\frac{dN_e}{dE_{\rm vis}}={}&
N_{\rm Ar}\int dE_\nu\int_{t_n^{\rm min}}^{t_n^{\rm max}}dt\;
\\
&{}\times F_{\nu_e}(E_\nu,t)\,
\sigma_{\rm ArCC}(E_\nu)\,W(E_{\rm vis},E_\nu),
\end{aligned}
\label{eq:dNdEvis-sm}
\end{equation}
and the time profile is
\begin{equation}
\begin{aligned}
\frac{dN_e}{dt}={}&
N_{\rm Ar}\int dE_\nu\int_0^{E_\nu}dE_{\rm vis}\;
\\
&{}\times F_{\nu_e}(E_\nu,t)\,
\sigma_{\rm ArCC}(E_\nu)\,W(E_{\rm vis},E_\nu).
\end{aligned}
\label{eq:dNdt-sm}
\end{equation}
Here $N_{\rm Ar}$ is the number of argon targets and $\sigma_{\rm ArCC}$ is the $\nu_e$--argon charged-current cross section as implemented in our numerical pipeline.
Contributions from excited ${}^{40}\text{K}^\ast$ states can be included via state-dependent thresholds, but they are subdominant for the burst-level comparisons made in the main text.

%%%%%%%%%%%%%%%%%%%%%%%%%%%%%%%%%%%%%%%%%%%%%%%%%%%%%%%%%%%%
\section{Robustness against scalar neutrino self-interaction feedback}
\label{sec:self-feedback}
%%%%%%%%%%%%%%%%%%%%%%%%%%%%%%%%%%%%%%%%%%%%%%%%%%%%%%%%%%%%

In the main text, the additional inner resonance is generated by the quark-sourced scalar correction to the \(\tau\tau\) neutrino mass entry. In this section we examine whether a scalar neutrino self-interaction, sourced by the \(\nu_\tau\) component produced during this conversion, can undo the adiabatic flavor evolution responsible for the neutronization-burst signature.

We focus on the direct refractive scalar neutrino self-interaction. In the contact-interaction treatment of scalar neutrino self-interactions, this contribution is nonzero for Majorana neutrinos and vanishes for Dirac neutrinos in the ultrarelativistic limit~\cite{Yang:2018yvk}. Restricting to a flavor-preserving \(\tau\tau\) coupling, the generated \(\nu_\tau\) component produces an additional diagonal contribution to the flavor-evolution Hamiltonian,
\begin{equation}
H_{\nu\nu}^{\rm refr}(r)
=
\mu(r)P_\tau(r)\,
{\rm diag}(0,0,1),
\label{eq:SM-Hnunu}
\end{equation}
where \(P_\tau(r)\) denotes the local \(\nu_\tau\) fraction of the neutronization-burst component,
\begin{equation}
P_\tau(r)
\equiv
\frac{n_{\nu_\tau}(r)}{n_{\rm burst}(r)},
\qquad
0\leq P_\tau(r)\leq1.
\label{eq:SM-Ptau}
\end{equation}
Before the additional conversion, the burst is predominantly \(\nu_e\) and \(P_\tau\simeq0\), so this term is initially absent and develops continuously as the conversion proceeds.

In the inner region where the additional SNSI-induced resonance occurs, the \(\nu_\mu\) state is approximately a spectator. We denote by \(\delta M_q(r)\equiv\delta M_{\tau\tau}(r)\) the quark-induced scalar mass entry already included in the main text. The relevant \(\nu_e\)--\(\nu_\tau\) Hamiltonian, now including the refractive feedback, is
\begin{multline}
H_{\rm refr}(r)=
\\
\begin{pmatrix}
V(r) & B_{\rm vac}+B_q(r) \\[1mm]
B_{\rm vac}^*+B_q^*(r) &
\dfrac{\left|\delta M_q(r)\right|^2}{2E}+\mu(r)P_\tau(r)
\end{pmatrix},
\label{eq:SM-Hrefr}
\end{multline}
where
\begin{equation}
B_q(r)
=
\frac{c_{e\tau}\delta M_q(r)}{2E},
\qquad
c_{e\tau}
=
\sum_j m_jU_{ej}U_{\tau j}^*,
\label{eq:SM-Bq}
\end{equation}
and
\begin{equation}
B_{\rm vac}
=
\frac{
\left[
U\,{\rm diag}(0,\Delta m_{21}^2,\Delta m_{31}^2)U^\dagger
\right]_{e\tau}
}{2E}.
\label{eq:SM-Bvac}
\end{equation}
Unlike the quark-induced SNSI term, the contribution in Eq.~\eqref{eq:SM-Hnunu} is added directly to the diagonal \(\tau\tau\) energy level. It does not modify \(\delta M_q(r)\) or the associated off-diagonal term \(B_q(r)\).

Let \(r_0\) denote the location of the additional SNSI-induced resonance in the absence of neutrino self-interaction feedback. Defining
\begin{equation}
x\equiv\frac{r}{r_0},
\qquad
V_0\equiv V(r_0),
\label{eq:SM-x}
\end{equation}
the original resonance condition is
\begin{equation}
\frac{\left|\delta M_q(r_0)\right|^2}{2E}
=
V_0.
\label{eq:SM-original-res}
\end{equation}
Using the inner-region density profile adopted in the main text, \(\rho(r)\propto r^{-3}\), one has
\begin{equation}
\begin{aligned}
V(r)&=V_0x^{-3}, \\
\delta M_q(r)&=\delta M_q(r_0)x^{-3}, \\
\frac{\left|\delta M_q(r)\right|^2}{2E}&=V_0x^{-6}, \\
B_q(r)&=B_0x^{-3},
\end{aligned}
\label{eq:SM-inner-scalings}
\end{equation}
where
\begin{equation}
B_0
\equiv
\frac{c_{e\tau}\delta M_q(r_0)}{2E}.
\label{eq:SM-B0}
\end{equation}

The refractive neutrino-neutrino contribution contains the angular factor \(1-\hat{\mathbf p}\cdot\hat{\mathbf q}\) associated with forward scattering between neutrino trajectories. As the burst neutrinos free stream outward, their local number density decreases as \(r^{-2}\). At radii larger than the neutrinosphere, their trajectories also become increasingly collinear with a radially propagating test neutrino, so the angular average of \(1-\hat{\mathbf p}\cdot\hat{\mathbf q}\) decreases approximately as \(r^{-2}\) in the bulb-geometry approximation. We therefore take
\begin{equation}
\mu(r)=\mu_0x^{-4}.
\label{eq:SM-mu-scaling}
\end{equation}

To compare this contribution with the quark-induced SNSI conversion studied in the main text, we define
\begin{equation}
\alpha\equiv\frac{\mu_0}{V_0}.
\label{eq:SM-alpha}
\end{equation}
Here \(\mu_0\) is the refractive \(\tau\tau\) potential that would be generated at the original resonance radius \(r_0\) by a fully converted local burst component. Since \(V_0=\left|\delta M_q(r_0)\right|^2/(2E)\), \(\alpha\) measures this refractive contribution relative to the quark-induced quadratic SNSI term that generates the original resonance.

Writing
\begin{equation}
\mu_0
\simeq
\frac{y_{\tau\tau}^{\,2}}{m_\phi^2}
D_0\,n_{\rm burst}(r_0),
\label{eq:SM-mu0}
\end{equation}
where \(D_0\) denotes the angular and normalization factor of the refractive source at \(r_0\), and using
\begin{equation}
\delta M_q(r_0)
=
\frac{y_{\tau\tau}y_q}{m_\phi^2}\,n_q(r_0),
\label{eq:SM-Mq-source}
\end{equation}
one obtains
\begin{equation}
\alpha
\simeq
\left|\frac{y_{\tau\tau}}{y_q}\right|
D_0\,
\frac{n_{\rm burst}(r_0)}{n_q(r_0)}
\frac{2E}{\left|\delta M_q(r_0)\right|}.
\label{eq:SM-alpha-couplings}
\end{equation}
For numerical illustration, we adopt the original no-feedback resonance radius
\begin{equation}
r_0=100\ {\rm km},
\end{equation}
corresponding to the inner-resonance benchmark used below. At this radius, the representative scales are
\begin{equation}
\frac{n_{\rm burst}(r_0)}{n_q(r_0)}
\sim
10^{-3},
\qquad
\frac{\left|\delta M_q(r_0)\right|}{E}
\sim
10^{-5}.
\end{equation}
Equation~\eqref{eq:SM-alpha-couplings} then gives
\begin{equation}
\alpha
\sim
2\times10^{2}D_0
\left|\frac{y_{\tau\tau}}{y_q}\right|.
\label{eq:SM-alpha-est}
\end{equation}
For the illustrative choice \(D_0=0.01\), representative of the angular suppression at \(r_0\sim100\ {\rm km}\), this becomes
\begin{equation}
\alpha
\sim
2
\left|\frac{y_{\tau\tau}}{y_q}\right|.
\label{eq:SM-alpha-yuk}
\end{equation}
Thus, the direct refractive contribution can compete with the quark-induced quadratic SNSI term that drives the original conversion when the neutrino and quark Yukawa couplings are of comparable magnitude, and it can dominate for moderate hierarchies between them.\footnote{A neutrino-induced correction to the scalar mass itself is additionally suppressed by the relativistic scalar bilinear. For the same illustrative scales, this correction becomes relevant only for \(\left|y_{\tau\tau}/y_q\right|\gtrsim10^9\).}

We now estimate how the refractive contribution changes the position and adiabaticity of the conversion. At the center of the conversion region, denoted by \(r_*=x_*r_0\), the two diagonal energy levels are equal and, in the two-level adiabatic approximation,
\begin{equation}
P_\tau(r_*)=\frac{1}{2}.
\label{eq:SM-equal-mixture}
\end{equation}
The resonance condition following from Eq.~\eqref{eq:SM-Hrefr} is therefore
\begin{equation}
V_0x_*^{-6}
+
\frac{\mu_0}{2}x_*^{-4}
=
V_0x_*^{-3}.
\label{eq:SM-refr-res}
\end{equation}
Equivalently,
\begin{equation}
x_*^3-1
=
\frac{\alpha}{2}x_*^2,
\qquad
r_*=x_*r_0.
\label{eq:SM-position}
\end{equation}
Since \(\alpha>0\), the right-hand side of Eq.~\eqref{eq:SM-position} is positive, so \(x_*>1\) and therefore \(r_*>r_0\). The center of the SNSI-induced conversion is consequently shifted to larger radius in the presence of scalar neutrino self-interactions.

The conversion remains adiabatic provided the avoided-crossing gap does not become small compared with the radial variation of the diagonal level splitting. Defining
\begin{equation}
\begin{aligned}
\Delta(r)&\equiv
\frac{\left|\delta M_q(r)\right|^2}{2E}
+\mu(r)P_\tau(r)-V(r), \\
B(r)&\equiv B_{\rm vac}+B_q(r),
\end{aligned}
\label{eq:SM-Delta-B}
\end{equation}
the two-level adiabaticity parameter at the equal-mixture point is
\begin{equation}
\gamma_{\rm refr}
=
\left.
\frac{4|B(r)|^2}{\left|d\Delta/dr\right|}
\right|_{r_*}.
\label{eq:SM-gamma-def}
\end{equation}
The derivative must be evaluated self-consistently because \(P_\tau(r)\) changes as the neutrinos convert. In the two-level approximation,
\begin{equation}
P_\tau(r)
=
\frac{1}{2}
\left[
1-
\frac{\Delta(r)}
{\sqrt{\Delta^2(r)+4|B(r)|^2}}
\right],
\label{eq:SM-Ptau-eigenstate}
\end{equation}
so that at \(r=r_*\),
\begin{equation}
\left.
\frac{\partial P_\tau}{\partial\Delta}
\right|_{r_*}
=
-\frac{1}{4|B(r_*)|}.
\label{eq:SM-Ptau-derivative}
\end{equation}
Introducing
\begin{equation}
\beta\equiv\frac{B_{\rm vac}}{B_0},
\qquad
\gamma_0\equiv\frac{4|B_0|^2r_0}{3V_0},
\label{eq:SM-beta-gamma0}
\end{equation}
where \(\gamma_0\) is the adiabaticity in the absence of refractive feedback within the inner-region approximation, differentiating Eq.~\eqref{eq:SM-Delta-B} gives
\begin{multline}
\gamma_{\rm refr}
=\gamma_0\frac{3x_*}{x_*^3+2}
\left|1+\beta x_*^3\right|^2
\\
\times\left[
1+\frac{V_0}{2|B_0|}
\frac{1-x_*^{-3}}{\left|1+\beta x_*^3\right|}
\right].
\label{eq:SM-gamma-refr}
\end{multline}
The term in square brackets results from the self-consistent build-up of the refractive contribution across the conversion region.

For the same representative inner-resonance benchmark, we use
\begin{equation}
\begin{gathered}
\gamma_0=2.16,\qquad \frac{V_0}{2|B_0|}=7.0\times10^3, \\
\beta=+3.0\times10^{-4}.
\end{gathered}
\label{eq:SM-benchmark}
\end{equation}
The choice \(\beta>0\) corresponds to the case in which \(B_{\rm vac}\) and \(B_q(r)\) add in the off-diagonal entry of the Hamiltonian. Equations~\eqref{eq:SM-position} and \eqref{eq:SM-gamma-refr} then give the results in Table~\ref{tab:SM-feedback-benchmark}.

\begin{table}[t]
\caption{Position and adiabaticity of the conversion for the inner-resonance benchmark with constructive off-diagonal contributions. The coupling ratio is evaluated for $D_0=0.01$.}
\label{tab:SM-feedback-benchmark}
\label{eq:SM-table}
\begin{ruledtabular}
\begin{tabular}{cccc}
$\alpha$ & $\left|y_{\tau\tau}/y_q\right|$ & $r_*~(\text{km})$ & $\gamma_{\rm refr}$ \\
\hline
0   & 0   & 100             & 2.16 \\
10  & 5   & $5.0\times10^2$ & $1.8\times10^3$ \\
100 & 50  & $5.0\times10^3$ & $7.0\times10^2$ \\
200 & 100 & $1.0\times10^4$ & $1.4\times10^3$ \\
\end{tabular}
\end{ruledtabular}
\end{table}

As the strength of the neutrino self-interaction is increased, the conversion is pushed progressively farther from its original location at \(r_0=100\ {\rm km}\). For example, for \(\alpha=10\) the center of the conversion occurs at \(r_*=5.0\times10^2\ {\rm km}\), while for \(\alpha=200\) it occurs at \(r_*=1.0\times10^4\ {\rm km}\). Nevertheless, the corresponding values of \(\gamma_{\rm refr}\) remain much larger than unity, showing that the conversion remains highly adiabatic in these examples. The conversion develops continuously over a finite radial interval. In particular, \(r_*\) is the point at which the evolving burst component has equal \(\nu_e\) and \(\nu_\tau\) fractions, while the \(\nu_\tau\) component and its associated self-interaction begin to develop at smaller radii. For strong feedback, this conversion region can extend over radial distances of order \(10^3\ {\rm km}\).

The numerical examples above correspond to the relative sign for which the scalar-induced and vacuum off-diagonal entries add. For the opposite relative sign, these entries cancel at the radius determined by
\begin{equation}
B_{\rm vac}+B_q(r)=0,
\qquad
x_{\rm cancel}=|\beta|^{-1/3}.
\end{equation}
For \(|\beta|=3.0\times10^{-4}\) and \(r_0=100\ {\rm km}\), one finds
\begin{equation}
x_{\rm cancel}\simeq15,
\qquad
r_{\rm cancel}\simeq1.5\times10^3\ {\rm km}.
\end{equation}
In the absence of neutrino self-interactions, this cancellation lies far outside the original conversion at \(r_0\) and does not affect it. The feedback can instead shift the conversion toward this radius. A suppression of adiabaticity is therefore expected only when the center of the displaced conversion lies close to \(r_{\rm cancel}\), which corresponds to \(\alpha\simeq30\) in the present benchmark. Away from this particular opposite-sign configuration, the neutrino self-interaction shifts the conversion outward without spoiling the adiabatic evolution found above.

For sufficiently large feedback, the conversion region can extend toward the ordinary MSW layers, where the inner \(2\times2\) approximation is no longer fully quantitative because vacuum diagonal terms and three-flavor mixing become important. Nevertheless, this does not by itself indicate a loss of adiabaticity. The refractive feedback mainly displaces the conversion to larger radii; the off-diagonal terms responsible for the avoided crossing remain present, while the matter profile becomes smoother at larger radius. We therefore see no generic reason for the adiabatic mapping to be spoiled when the conversion is shifted outward, although a complete description of the detailed flavor history in this regime would require solving the full three-flavor evolution with the self-consistent refractive term. The key point for the signal discussed in the main text is that, if the evolution remains adiabatic, the initially populated matter eigenstate is still mapped to the same vacuum state outside the star, and the neutronization-burst signature is preserved.

Finally, the scalar neutrino self-interaction feedback does not change the minimum value of \(\left|\eta_{\tau\tau}^{\odot}\right|\) required to trigger the effect in the supernova environment. This threshold is set by the condition that the quark-induced quadratic SNSI term dominates over the standard matter potential near the neutrinosphere and then crosses it at larger radius, generating the additional SNSI-induced conversion. The feedback analyzed here acts only after that conversion has begun to produce a local \(\nu_\tau\) component, and therefore displaces a conversion already generated by the quark-induced SNSI term rather than creating the original resonance. Thus the threshold remains \(\left|\eta_{\tau\tau}^{\odot}\right|\sim10^{-4}\), as in the main analysis.

%%%%%%%%%%%%%%%%%%%%%%%%%%%%%%%%%%%%%%%%%%%%%%%%%%%%%%%%%%%%
\section{Shape-only comparison of the DUNE time profile}
\label{sec:shape-only-chi2}
%%%%%%%%%%%%%%%%%%%%%%%%%%%%%%%%%%%%%%%%%%%%%%%%%%%%%%%%%%%%

In the main text, we compare the DUNE time distributions associated with the standard MSW case and with the SNSI-induced flavor conversion. Since uncertainties in the \(\nu_e\)--\(^{40}{\rm Ar}\) cross section and in the overall supernova flux normalization can affect the absolute number of events, we quantify here the separation between the time profiles after fully marginalizing over the normalization.

We use the binned time distributions in the left panel of Fig.~\ref{fig:NO positive} of the main text. The standard MSW prediction is treated as the reference spectrum, while the fully adiabatic and partially adiabatic SNSI cases are compared against it. For this purpose, we use the Poisson-likelihood estimator
\begin{equation}
\begin{aligned}
\chi^2=\min_{\xi}\sum_i2\biggl[&
(1+\xi)F_i-D_i
\\
&{}+D_i\ln\left(\frac{D_i}{(1+\xi)F_i}\right)
\biggr],
\end{aligned}
\label{eq:SM-shape-chi2}
\end{equation}
where \(D_i\) is the standard MSW prediction in the \(i\)-th time bin and \(F_i\) is the corresponding SNSI prediction. The nuisance parameter \(\xi\) is left completely free, subject only to \(1+\xi>0\), so that Eq.~\eqref{eq:SM-shape-chi2} measures only the difference in the time-profile shape.

Equivalently, writing \(a=1+\xi\), the minimum occurs at
\begin{equation}
a_{\rm bf}
=
\frac{\sum_i D_i}{\sum_i F_i}.
\label{eq:SM-best-fit-normalization}
\end{equation}
This removes the effect of an overall rescaling of the SNSI prediction and isolates the time dependence of the signal.

Using the binned event rates shown in Fig.~\ref{fig:NO positive} of the main text, we find the results in Table~\ref{tab:SM-shape-chi2}.

\begin{table}[t]
\caption{Shape-only comparison of the SNSI time profiles with the standard MSW reference spectrum. The normalization $a_{\rm bf}$ is fitted freely.}
\label{tab:SM-shape-chi2}
\begin{ruledtabular}
\begin{tabular}{ccc}
Case & $a_{\rm bf}$ & $\chi^2_{\rm min}$ \\
\hline
$\gamma\gg1$       & 0.60 & 37.8 \\
$\gamma\lesssim1$ & 0.74 & 16.5 \\
\end{tabular}
\end{ruledtabular}
\end{table}

The dominant contribution to \(\chi^2\) comes from the \(5\)--\(10\,{\rm ms}\) bin, where the SNSI cases predict a restored neutronization-burst peak while the standard MSW prediction remains suppressed. Thus, even after the overall normalization is allowed to float freely, the time-profile shape remains strongly distinguishable.

\end{document}